\documentclass[a4paper,english,numberwithinsect]{llncs}
\usepackage{hyperref}
\usepackage{amssymb}
\usepackage[ruled, vlined, linesnumbered]{algorithm2e}
\usepackage{todonotes}
\usepackage{mathtools}
\usepackage{wasysym}
\usepackage{stmaryrd}
\usepackage{graphicx}
\usepackage[font=small,labelfont=bf]{caption}
\usepackage[font=small,labelfont=normal]{subcaption}
\usepackage{multirow}
\usepackage{paralist}
\usepackage{cleveref}

\newcommand{\caux}{c_{\mathrm{aux}}}
\newcommand{\cauxx}[2]{\caux(#1, #2)}
\newcommand{\col}{\mathrm{col}}
\newcommand{\NH}{\mathcal{N}}

\newcommand{\level}[3]{\mathrm{level}_{#1}^{#2}(#3)}

\DeclareMathSymbol{\shortminus}{\mathbin}{AMSa}{"39}
\newcommand{\tailtriangle}{\kern-0.1em\shortminus\kern-0.1em\triangleleft}

\title{On $3$-Coloring Circle Graphs\thanks{Funded by the Deutsche Forschungsgemeinschaft (German Research
Foundation, DFG) under grant RU-1903/3-1.}}
\author{Patricia Bachmann\orcidID{0009-0003-3749-6265} \and Ignaz Rutter\orcidID{0000-0002-3794-4406}
\and Peter Stumpf\orcidID{0000-0003 -0531 -9769}}
\institute{
    Faculty of Informatics and Mathematics, University of Passau, Germany
    \email{\{bachmanp,rutter,stumpf\}@fim.uni-passau.de}
}

\begin{document}\pagestyle{plain}

\maketitle\thispagestyle{plain}

    \begin{abstract}
        Given a graph $G$ with a fixed vertex order $\prec$, one obtains a circle graph $H$ whose vertices are the
        edges of $G$ and where two such edges are adjacent if and only if their endpoints are pairwise distinct and
        alternate in $\prec$.
        Therefore, the problem of determining whether $G$ has a $k$-page book embedding with spine order~$\prec$ is
        equivalent to deciding whether $H$ can be colored with $k$ colors.
        Finding a $k$-coloring for a circle graph is known to be NP-complete for $k \geq 4$ and trivial for $k \leq 2$.
        For~$k = 3$, Unger (1992) claims an efficient algorithm that finds a 3-coloring in $O(n \log n)$ time, if it
        exists.
        Given a circle graph $H$, Unger's algorithm (1) constructs a 3-\textsc{Sat} formula~$\Phi$ that is
        satisfiable if and only if $H$ admits a 3-coloring and (2) solves~$\Phi$ by a backtracking strategy that
        relies on the structure imposed by the circle graph.
        However, the extended abstract misses several details and Unger refers to his PhD thesis (in German) for
        details.

        In this paper we argue that Unger's algorithm for 3-coloring circle graphs is not correct and that
        3-coloring circle graphs should be considered as an open problem.
        We show that step (1) of Unger's algorithm is incorrect by exhibiting a circle graph whose formula $\Phi$ is
        satisfiable but that is not 3-colorable.
        We further show that Unger's backtracking strategy for solving
        $\Phi$ in step (2) may produce incorrect results and give empirical evidence that
        it exhibits a runtime behaviour that is not consistent with the claimed running time.
    \end{abstract}

    \section{Introduction}
    \label{sec:intro}

    Let $G = (V,E)$ be a graph.
    A \emph{$k$-page book embedding} of $G$ is a total order $\prec$ of~$V$ and a partition of $E$ into $k$ sets~$E_1
    ,\dots,E_k$, called \emph{pages} such that no page~$E_i$ contains two edges~$\{u_1,v_1\}$, $\{u_2,v_2\}$ with~$v_1 \prec v_2 \prec u_1 \prec u_2$. The \emph{page number} (also called \emph{stack number}) of a graph is the smallest~$k$ such that $G$ admits a $k$-page booking embedding.

    Book embeddings are a central element to graph drawing.
    They have been studied in the context of VSLI design~\cite{Chung1987}, arc diagrams~\cite{Wattenberg} and circular
    layouts~\cite{Baur2004} as well as clustered planarity~\cite{Hong2018} and simultaneous embedding~\cite{Angelini2012}.
    There is a plethora of results that aim at bounding the page numbers of various graph classes. For example, the page
    number of planar graphs is~4~\cite{DBLP:journals/jocg/KaufmannBKPRU20,DBLP:journals/jcss/Yannakakis89} and the
    page number of 1-planar graphs is at most 39~\cite{DBLP:conf/esa/BekosB0R15}.

    Since computing the page number is NP-complete~\cite{chung1985graph}, often additional restrictions are imposed. One such restriction is to find a $k$-page book embedding with a fixed order~$\prec$.
    This problem is closely related to the $k$-coloring problem on circle graphs.
    A \emph{$k$-coloring} of a graph $G$ is a function$~{\mathrm{col}: V(G) \to \{1, \dots, k\}}$ such that $\col (u) \neq \col (v)$ for every~$\{u,v\} \in E(G)$.
    A \emph{circle graph} is an undirected graph $H$ that has an intersection representation with chords of a circle. More precisely, in a \emph{chord diagram} of $H$ we represent each vertex $v \in V(H)$ by a chord~$C_v$ such that two chords $C_u$ and~$C_v$ intersect if and only if~$\{u,v\} \in E(H)$.

\begin{figure}[tb]
        \centering
        \includegraphics{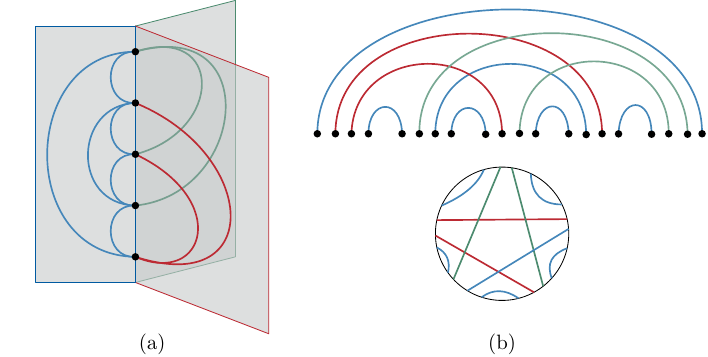}
        \caption{(a) Book embedding for $K_5$. (b) Chord diagram of the corresponding circle graph~$H$.}
        \label{fig:book-embedding}
    \end{figure}

    Finding a $k$-page book embedding of a graph $G=(V,E)$ with a
    fixed vertex order~$\prec$ is equivalent to solving the
    $k$-coloring problem for circle graphs. To see this, construct a
    graph $H$ such that $V(H) = E$ and two edges in $V(H)$ are
    adjacent in $H$ if and only if their endpoints alternate
    in~$\prec$. Then two edges of $G$ can be in the same page of a
    book embedding with order~$\prec$ if and only if they are not
    adjacent in $H$. Therefore we obtain a bijection between the
    $k$-page book embeddings with order~$\prec$ and the $k$-colorings
    of $H$. It is further readily seen that $H$ admits a chord
    diagram; see Fig.~\ref{fig:book-embedding} for an illustration.

    The $k$-coloring problem for circle graphs is known to be NP-complete for $k \geq 4$~\cite{Unger} and efficiently solvable
    for $k \leq 2$.
    The case $k = 3$ remained as an open problem until Unger claimed it to be solvable in polynomial time~\cite{Unger1992}.
    Unfortunately, in the publication many details and proofs are missing and no journal version followed.
    Instead, Unger refers to his PhD thesis~\cite{DBLP:books/daglib/0070152}
    for a full version, which is written in German and not available online. This bad state of affairs has been pointed
    out by David Eppstein in a
    blog post, where he writes that the problem should be
    considered open~\cite{blogpost}, as well as by Dujmović and Wood~\cite{Dujmovic2004}.

    In this paper we present Unger's ideas for an efficient algorithm for the $3$-coloring problem on circle graphs
    and show, using both counterexamples and empirical results, why $3$-coloring circle graphs and therefore
    the $3$-page book embedding problem should indeed be considered open problems.

    Throughout this work let~$G=(V,E)$ be a circle graph, for which we
    want to decide the existence of a $3$-coloring. We assume without
    loss of generality that $G$ is connected and contains no
    induced~$K_4$.

    \section{Unger's 3-Coloring Algorithm}
    \label{sec:unger}

    For a graph~$H$
    let~$\mathcal P(H) = \{ \{u,v\} \in \binom{V(H)}{2} \setminus E(H)
    \mid N(u) \cap N(v) \ne \emptyset\}$ be the pairs of non-adjacent
    vertices that have a common neighbor and
    let~$\mathcal D \subseteq \mathcal P(H)$. An \emph{auxiliary
    coloring function for~$\mathcal D$} is a function
    $\caux \colon \mathcal D \to \{\mathtt{true},\mathtt{false}\}$.
    An auxiliary coloring function is \emph{realizable} if there
    exists a 3-coloring~$\mathrm{col}$ of $H$ such that for each
    pair~$\{x,y\} \in \mathcal D$ we
    have~$\col (x) = \col (y) \iff \caux(\{x,y\}) = \mathtt{true}$.

    Let now~$G=(V,E)$ be a graph and let~$\mathcal G$ be a set of
    induced subgraphs of $G$ and
    let~$\mathcal P(\mathcal G) = \bigcup_{H \in \mathcal G} \mathcal P(H)$.
    An \emph{auxiliary coloring function of $G$ with respect
    to~$\mathcal G$} is an auxiliary coloring function
    for~$\mathcal P(\mathcal{G})$. Such a function is called \emph{consistent}
    if its restriction $\caux|_{\mathcal P(H)}$ is realizable for
    each~$H \in \mathcal G$.

    In his approach, Unger constructs for a given circle graph~$G$ a
    family~$\mathcal G$ of induced subgraphs, which he calls important
    subgraphs, such that
    \begin{inparaenum}[(i)]
        \item  an auxiliary coloring function~$\caux$ with respect to~$\mathcal G$ is consistent if and only if it is realizable,
        \item  the existence of a consistent auxiliary coloring function can be expressed by a 3-SAT formula~$\Phi$ that
        \item can be solved efficiently with a backtracking algorithm.
    \end{inparaenum}

\begin{figure}[tb]
        \centering
        \includegraphics{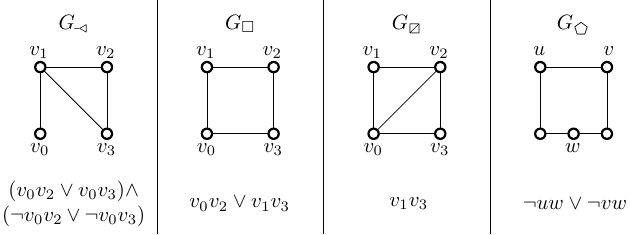}
        \caption{Important subgraphs and the clauses they contribute to $\Phi$.}
        \label{fig:imp-subgraphs}
    \end{figure}

    Let~$\mathcal G'$ be the family of induced subgraphs of $G$ that are
    isomorphic to one of the graphs~$G_{\tailtriangle}$,
    $G_{\square}$, $G_{\pentagon}$, $G_{\boxslash}$ from Fig.~\ref{fig:imp-subgraphs}
    or to a cycle~$C_k$ with $k\ge 6$. For each $H \in \mathcal G'$
    there is a formula~$\Phi(H)$ whose satisfying truth assignments
    are the realizable auxiliary coloring functions for $H$. For the
    three graphs~$G_{\tailtriangle}$, $G_{\square}$ and
    $G_{\boxslash}$, the corresponding formula~$\Phi(H)$ is in fact a
    2-SAT formula as shown in Fig.~\ref{fig:imp-subgraphs}, where for $C_k$ with~$k \ge 5$, the
    formula~$\Phi(H)$ given by Unger is a 3-SAT formula of size linear
    in~$k$, which uses some additional variables; see Appendix~\ref{sec:bt-counter} for details.
    For $G_{\pentagon}$, i.e.\ cycles with $k = 5$, Unger additionally uses the 2-SAT clauses
    shown in Fig.~\ref{fig:imp-subgraphs}.
    Then the existence of a consistent auxiliary coloring function for~$\mathcal G'$ can be expressed by the formula~$\Phi(\mathcal G') = \bigwedge_{H \in \mathcal G'}\Phi(H)$.

\begin{figure}[tb]
        \centering
        \includegraphics{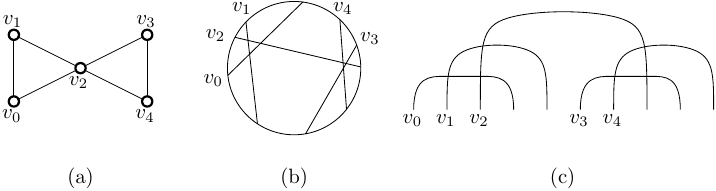}
        \caption{(a) An undirected graph $G$. (b) Representation of $G$ of chords on a circle. (c) Alternative circle
        graph representation for $G$.}
        \label{fig:circle-graph}
    \end{figure}

    The family~$\mathcal G'$ is however, still too large, e.g., it may contain an exponential number of cycles. Therefore Unger restricts his important subgraphs~$\mathcal G$ to a subset of~$\mathcal G'$ that is defined according to a chord diagram of $G$. To this end, take a
    chord diagram of $G$ and consider it to be cut open and rolled out
    such that the chords form arcs over a straight line; see
    Fig.~\ref{fig:circle-graph}(c). In what follows we identify each vertex~$v$ with its chord~$C_v$.

\begin{figure}[tb]
        \centering
        \includegraphics{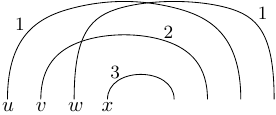}
        \caption{Chord $u$ encases $v$ and $x$ and directly encases $v$; $v$ and $w$ directly encase~$x$.}
        \label{fig:encase}
    \end{figure}

    A chord~$u$ \emph{encases} a chord~$v$ if the endpoints of~$v$
    lie between the endpoints of~$u$. Further~$u$ \emph{directly
    encases} $v$ if it encases~$v$ and there is no
    vertex~$w \ne u,v$ such that $u$ encases~$w$ and $w$
    encases~$v$; see Fig.~\ref{fig:encase} for an example.

    The \emph{levels} of a circle graph $G$ are recursively defined as
    \[\level{G}{}{l} = \begin{cases}
                           \{v \in V(G) \colon \nexists u \in V(G) \colon u \text{ encases } v\} & l=1\\
                           \{v \in V(G) \colon \exists u \in \level{G}{}{l-1} \colon u \text{ directly
                           encases } v\} &  l > 1\\
    \end{cases}\,.\]

    The set of important subgraphs~$\mathcal G$ consists of those
    graphs~$H \in \mathcal G'$ where if $H$ is isomorphic
    to~$G_{\pentagon}$ or to $C_k$ with~$k\ge 6$, then all vertices of $H$
    belong to two adjacent levels, and otherwise for each
    pair~$\{u,v\} \in \mathcal P(H)$ the vertices $u$ and $v$ are either
    on the same level or one directly encases the other. Unger's
    algorithm relies on two claims~\cite[p.394 ff.]{Unger1992}: (1) The graph $G$ is 3-colorable if
    and only if there exists a consistent auxiliary coloring function
    with respect to~$\mathcal G$.  (2) The formula~$\Phi(\mathcal G)$
    can be solved efficiently by a backtracking algorithm whose search
    tree has $O(\log n)$ leaves.

    \section{The Counterexample}
    \label{sec:counterex}

\begin{figure}[tb]
        \centering
        \includegraphics[width=\textwidth]{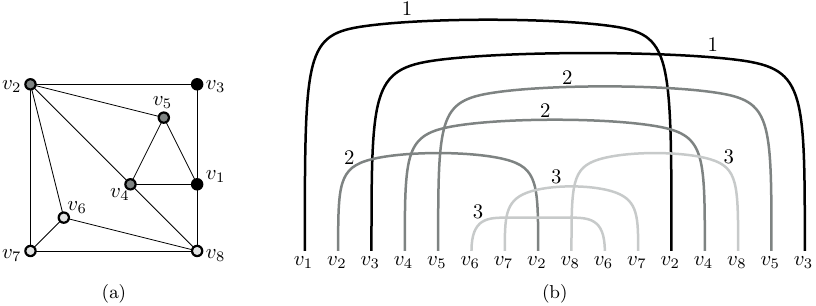}
        \caption{Counterexample $G$. Lighter shade indicate higher level.}
        \label{fig:counterexample}
    \end{figure}

    We show that Unger's claim (1) is false by giving a
    counterexample. Let~$G$ be the graph given in
    Fig.~\ref{fig:counterexample}a. Observe that $G$ is a circle
    graph as witnessed by the chord diagram in Fig.~\ref{fig:counterexample}b.

    We first show that $G$ is not $3$-colorable. Namely, the two subgraphs
    induced by~$\{v_1,v_2,v_4,v_5\}$ and $\{v_2,v_6,v_7,v_8\}$ imply~$\col(v_1) = \col (v_2)$ and
    $\col (v_2) = \col (v_8)$, respectively. However $v_1$ and~$v_8$ are adjacent.

    On the other hand, we show that for the family~$\mathcal G$ of
    important subgraphs of $G$ with respect to the chord diagram in Fig.~\ref{fig:counterexample},
    the formula~$\Phi(\mathcal G)$ is
    satisfiable. We first give the important subgraphs of $G$ and
    then construct the corresponding formula. In Fig.~\ref{fig:counterexample}a the
    vertices are colored according to their levels, where lighter
    colors indicate higher levels. It is not hard to see that $G$
    contains no induced cycle isomorphic to~$G_{\pentagon}$ or to~$C_k$
    with~$k \ge 6$ that is contained in two adjacent levels. It hence
    suffices to find the important subgraphs isomorphic
    to~$G_{\tailtriangle}$,$G_{\square}$, and $G_{\boxslash}$.

    We start with~$G_{\square}$ and~$G_{\boxslash}$. Observe that
    each of the pairs~$\{v_2,v_8\}$ and~$\{v_3,v_8\}$ neither lies on
    the same level nor does one the vertices directly encase the
    other. Hence no important subgraph contains both vertices of
    these pairs. For~$G_{\square}$ this leaves only the subgraphs
    listed in Table~\ref{tab:imp-sg}a and, similarly, for~$G_{\boxslash}$ only the
    subgraphs listed in Table~\ref{tab:imp-sg}b. Finally, it is straightforward to check that for~$G_{\tailtriangle}$ the
    only subgraphs are listed in Table~\ref{tab:imp-sg}c.
    A detailed description of all important subgraphs in $G$ is given in Appendix~\ref{subsec:important-subgraphs}.

    Table~\ref{tab:imp-sg} shows the clauses of the
    formula~$\Phi := \Phi(\mathcal G)$. Finally,
    Table~\ref{tab:assignment} gives a satisfying truth assignment for
    $\Phi$.
    The underlined literals in Table~\ref{tab:imp-sg} are those satisfied by that truth assignment.
    Since every clause contains a satisfied literal,~$\Phi$ is satisfiable and the truth assignment
    defines a consistent auxiliary coloring function for~$\mathcal G$.
    However, as $G$ is not 3-colorable, this contradicts Unger's
    claim~(1).

    \begin{table}[tb]
        \caption{Important subgraphs and the corresponding clauses for each of~(a)~$G_{\square}$, (b) $G_{\boxslash}$, and (c) $G_{\tailtriangle}$. For the satisfying variable assignment from Table~\ref{tab:assignment} all true literals are underlined.}

        \begin{center}
            \begin{subtable}{.45\textwidth}
                \centering
                \begin{tabular}{ | c | c | }
                    \hline
                    Important Subgraph $G_\square$ & Clauses                                          \\
                    \hline
                    $\{v_1, v_2, v_3, v_4\}$       & \underline{$v_1 v_2$} $\vee \underline{v_3 v_4}$ \\
                    \hline
                    $\{v_1, v_2, v_3, v_5\}$       & \underline{$v_1 v_2$} $\vee v_3 v_5$             \\
                    \hline
                \end{tabular}
                \caption{}
            \end{subtable}
            \begin{subtable}{.45\textwidth}
                \centering
                \begin{tabular}{ | c | c | }
                    \hline
                    Important Subgraph $G_{\boxslash}$ & Clauses              \\
                    \hline
                    $\{v_1, v_2, v_4, v_5\}$           & \underline{$v_1v_2$} \\
                    \hline
                    $\{v_1, v_4, v_5, v_8\}$           & $\underline{v_5v_8}$ \\
                    \hline
                \end{tabular}
                \caption{}
            \end{subtable}
            \begin{subtable}{\textwidth}
                \centering
                \begin{tabular}{ | c | c | }
                    \hline
                    Important Subgraph $G_{\tailtriangle}$ & Clauses                                                                              \\
                    \hline
                    $\{v_1, v_6, v_7, v_8\}$ & $(v_1v_6 \vee \underline{v_1v_7}) \wedge (\underline{\neg
                    v_1v_6} \vee \neg$ $v_1v_7)$ \\
                    \hline
                    $\{v_1, v_3, v_4, v_5\}$               & $(\underline{v_3v_4} \vee v_3v_5) \wedge (\neg v_3v_4 \vee \underline{\neg v_3v_5})$ \\
                    \hline
                    $\{v_2, v_3, v_4, v_5\}$               & $(\underline{v_3v_4} \vee v_3v_5) \wedge (\neg v_3v_4 \vee \underline{\neg v_3v_5})$ \\
                    \hline
                    $\{v_2, v_4, v_6, v_7\}$               & $(\underline{v_4v_6} \vee v_4v_7) \wedge (\neg v_4v_6 \vee \underline{\neg v_4v_7})$ \\
                    \hline
                    $\{v_2,  v_5, v_6, v_7\}$                & $(v_5v_6 \vee \underline{v_5v_7}) \wedge (\underline{\neg v_5v_6} \vee \neg v_5v_7)$ \\
                    \hline
                    $\{v_2, v_4, v_5, v_6\}$               & $(\underline{v_4v_6} \vee v_5v_6) \wedge (\neg v_4v_6 \vee \underline{\neg v_5v_6})$ \\
                    \hline
                    $\{v_2, v_4, v_5, v_7\}$               & $(v_4v_7 \vee \underline{v_5v_7}) \wedge (\underline{\neg v_4v_7} \vee \neg v_5v_7)$ \\
                    \hline
                \end{tabular}
                \caption{}
            \end{subtable}
            \label{tab:imp-sg}
        \end{center}
    \end{table}
    \begin{table}[tb]
        \caption{Satisfying variable assignment for $\Phi$.}
        \begin{center}
            \begin{tabular}{ | c | c | }
                \hline
                True  & $v_1v_2$, $v_1v_7$, $v_2v_5$, $v_3v_4$, $v_4v_6$, $v_5v_7$, $v_5v_8$ \\
                \hline
                False & $v_1v_6$, $v_4v_7$, $v_5v_6$, $v_3v_5$                               \\
                \hline
            \end{tabular}
            \label{tab:assignment}
        \end{center}
    \end{table}

    \paragraph{Different Notions of Important Subgraphs.}
    \label{sec:ext}

    We note that the definition of important subgraphs subtly differs
    between the extended abstract~\cite{Unger1992} and Unger's PhD
    thesis~\cite{DBLP:books/daglib/0070152}. Namely, in the thesis,
    important subgraphs are not defined only via direct encasing but for some types
    their vertices also have to belong to (at most) two adjacent
    levels. The counterexample above refutes the claim from the
    extended abstract. The example in
    Fig.~\ref{fig:counterexample-diss} in the appendix refutes the analogous claim
    from Unger's thesis; see Appendix~\ref{sec:second-counter} for
    details.

    \section{Unger's Backtracking Algorithm}
    \label{sec:experiments}

    In addition to the counterexample, we investigated the backtracking algorithm described by Unger.
    Let $\Phi_1$ be the 2-\textsc{Sat} instance obtained from the important subgraphs
    $G_{\tailtriangle}$, $G_{\square}$, $G_{\boxslash}$ and $G_{\pentagon}$.
    Checking if $\Phi_1$ is solvable can be done in polynomial time.
    For the SAT instance $\Phi_2$ obtained from the remaining important subgraphs $G_{\pentagon}$
    and $C_k$ for $k \geq 6$ it is less clear how to solve it efficiently.
    Unger~\cite{Unger1992} proposes the following modified backtracking algorithm.
    When encountering a clause during the variable assignment whose literals are all set to
    \texttt{False}, we may jump in the backtracking tree to a
    variable which (1) results in a literal of that clause being set to \texttt{True} by flipping
    its assigned value and (2) the child that corresponds to this new assigned value is not included in the
    backtracking tree yet.
    We note that we rephrased property (2) from ``[t]he new value of [the variable] is not already
    included in the backtracking tree''~\cite[p. 396]{Unger1992}.
    If no such variable exists, the algorithm terminates and reports that there is no solution for $\Phi$.
    If a solution is found, it is used to compute a 3-coloring.

\begin{figure}
        \centering
        \begin{subfigure}{0.49\textwidth}
            \includegraphics[width=\textwidth]{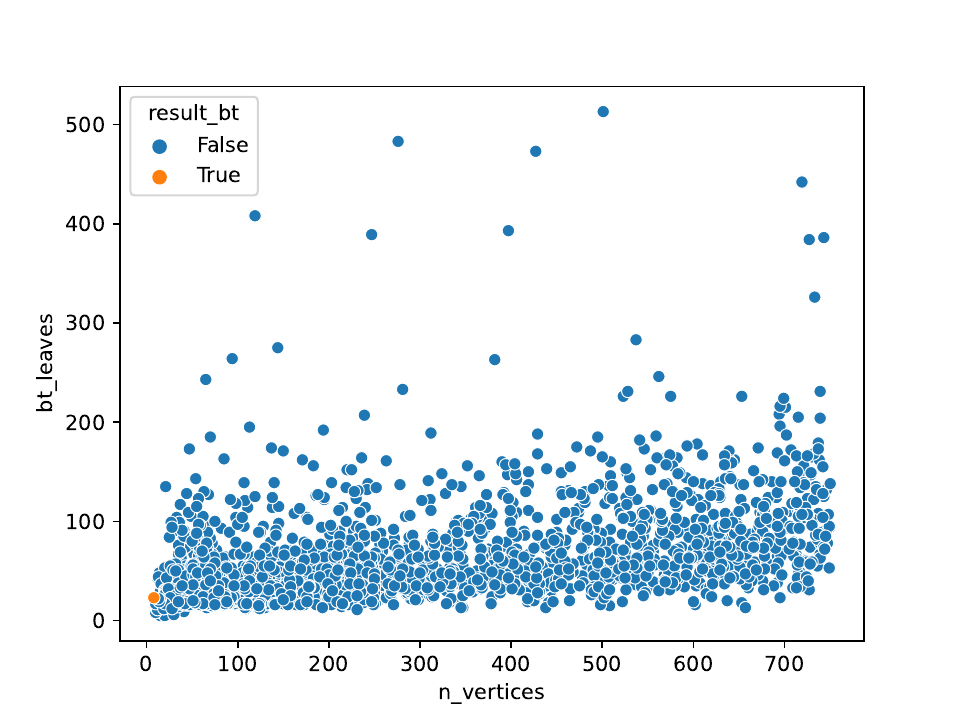}
            \caption{}
        \end{subfigure}
        \begin{subfigure}{0.49\textwidth}
            \includegraphics[width=\textwidth]{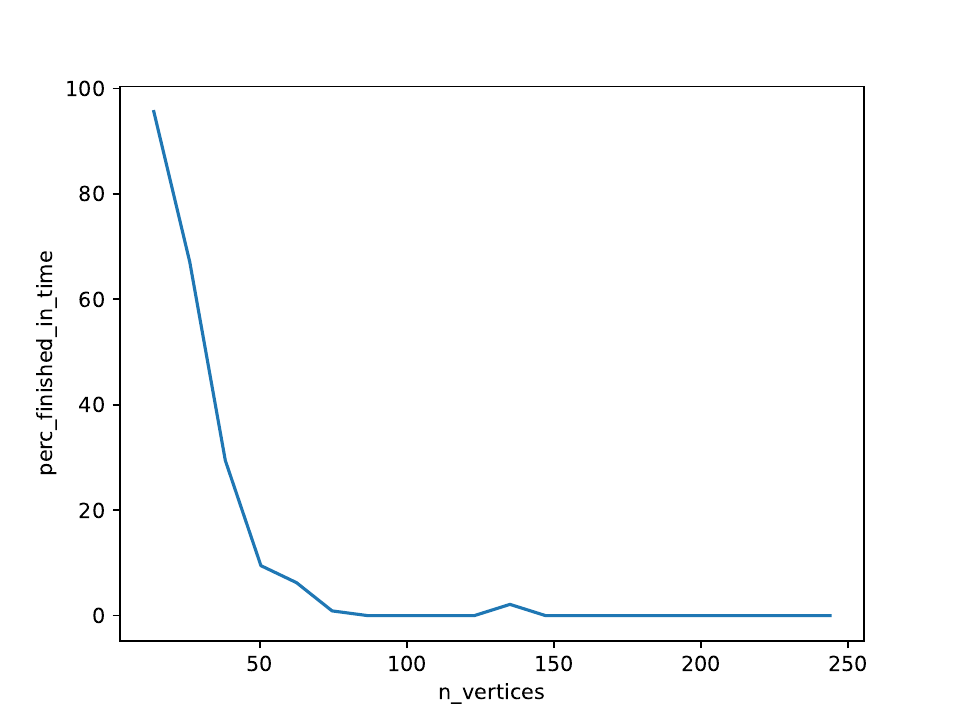}
            \caption{}
        \end{subfigure}
        \caption{(a) Number of leaves in Unger's backtracking tree.
            (b) Percentage of instances solved by regular backtracking within a one-hour time limit.
        }
        \label{fig:eval}
    \end{figure}

    Unger claims that this backtracking tree has at most $O(\log n)$ leaves.
    To verify this claim we implemented his backtracking algorithm, ran it on randomly generated
    3-colorable circle graphs and counted the number of leaves in the backtracking tree; see Fig.~\ref{fig:eval}(a)
    and Appendix~\ref{sec:test-data} for further details on the test data.
    While the algorithm is reasonably fast, the number of leaves scatters quite a bit, e.g.\ some graphs
    produce backtracking trees with around 500 leaves even though they have only 250 to 500 vertices.
    This is likely not consistent with an upper bound of $O(\log n)$.
    More importantly, while all the graphs are 3-colorable, the backtracking algorithm reports no solution
    for all but one of the instances.
    In Appendix~\ref{sec:bt-counter}, we show how the algorithm may already fail for a cycle of length~5
    if the order of variable assignments is poorly chosen.

    Therefore, we also evaluated a regular backtracking algorithm on the same instances.
    We found that the number of leaves of these backtracking trees grows exponentially.
    Fig.~\ref{fig:eval}(b) shows the percentage of solved instances within a time limit of one hour.
    Notably, starting at around 50 vertices, the algorithm barely manages to solve any instances.
    For readability, the Figure shows only the fraction of solved instances with up to 250 vertices as the backtracking
    algorithm does not terminate on any of the larger instances.
    This indicates the impracticality of a regular backtracking approach and that crucial insights are still missing to
    see how a backtracking approach could be modified to be more efficient on circle graphs.

    Finally, we also implemented the coloring algorithm that uses the solution for $\Phi$ to color the graph.
    Our result is that less than 20\% of the computed colorings were valid.
    This shows that claim (1) not only fails qualitatively du to the counterexample from Section~\ref{sec:counterex}
    but also quantitatively for the vast majority of instances.

    \section{Conclusion}
    \label{sec:conclusion}
    We have shown that the question of whether 3-coloring for circle graphs is
    possible in polynomial time should be considered open, even though Unger
    claimed to provide a polynomial time algorithm in a conference paper in
    1992~\cite{Unger1992}. To this end, we provided two counterexamples: one that contradicts the characterization
    in terms of the auxiliary coloring function and one that shows that the modified backtracking algorithm
    may fail to compute a correct solution.
    We further gave empirical evidence that displays a discrepancy to the claimed running time.

    Considering the approach with important subgraphs, it appears that
    especially large induced cycles increase the difficulty, since for the other
    important subgraphs only a 2-\textsc{Sat} formula is constructed.

    \begin{question}
        What is the complexity of 3-coloring for circle graphs where no $C_k$
        with $k>4$ is an induced subgraph?
    \end{question}

    We note that when considering all important subgraphs containing four vertices for the 2-\textsc{Sat}
    instance disregarding levels, we were not able to find a counterexample similar to the one in
    Section~\ref{sec:counterex}.

    \bibliographystyle{splncs04}
    \bibliography{references}

    \appendix
    \newpage

    \section{Important Subgraphs of the Counterexample}
    \label{subsec:important-subgraphs}

    We show for our main counterexample that we can construct a satisfiable formula $\Phi$.
    For this, we first find all important subgraphs contained in $G$.
    We know that only those subgraphs are relevant to us that contribute clauses to $\Phi$, i.e.\ clauses that
    contain values for pairs of vertices for which $\caux$ is defined.
    Therefore, the non-crossing chords in an important subgraph either lie within the same level or one set of chords
    directly encases another.

    \subsection{Important Subgraph $G_{\tailtriangle}$}
    \label{subsubsec:imp1}

    We first consider chords within one level.
    Since every level contains less than four vertices each, see Fig.~\ref{fig:counterexample},
    this graph contains no $G_{\tailtriangle}$ with all chords on the same level.

\begin{figure}[tb]
        \centering
        \includegraphics[width=\textwidth]{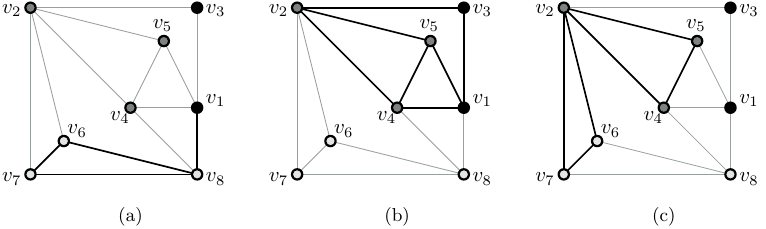}
        \caption{The different induced $G_{\tailtriangle}$}
        \label{fig:counterex-tailtriangle}
    \end{figure}

    Next, we consider $G_{\tailtriangle}$ with directly encased chords.
    For $\level{G}{}{1}$ and we have that $v_1$ directly encases $\{v_2, v_6, v_7\}$ while $v_3$ directly
    encases $\{v_4, v_5\}$.
    We have $\NH_G(v_1) \cap \NH_G(v_2) = \{v_3, v_4, v_5\} \eqqcolon \NH_{G|\{v_1, v_2\}}$ and since there is no
    3-clique containing $v_2$ and two chords of $\NH_{G|\{v_1, v_2\}}$ such that $v_1$ crosses only of the chords of
    the 3-clique, there is no induced $G_{\tailtriangle}$ containing $v_1$ and $v_2$.
    Further, we have $\NH_G(v_1) \cap \NH_G(v_6) \cap \NH_G(v_7) = \{v_8\}$, so $\{v_1, v_6, v_7, v_8\}$ induces a
    $G_{\tailtriangle}$, see Fig.~\ref{fig:counterex-tailtriangle}(a).
    For $v_3$ we have that $\NH_G(v_3) \cap \NH_G(v_4) \cap \NH_G(v_5) = \{v_1, v_2\}$, therefore $\{v_1, v_3, v_4,
    v_5\}$ and $\{v_2, v_3, v_4, v_5\}$ each induce a $G_{\tailtriangle}$ conveying the same information for our 2-\textsc{Sat}
    instance, see Fig.~\ref{fig:counterex-tailtriangle}(b).

    For $\level{G}{}{2}$ we have that $v_2$ encases no chords, $v_4$ directly encases $\{v_6, v_7\}$
    and $v_5$ directly encases $\{v_6, v_7, v_8\}$.
    We have that $\NH_G(v_4) \cap \NH_G(v_5) \cap \NH_G(v_6) \cap \NH_G(v_7) = \{v_2\}$, therefore $\{v_2, v_4, v_5,
    v_6, v_7\}$ induces a $G_{\tailtriangle}$, see Fig.~\ref{fig:counterex-tailtriangle}(c).
    For $v_4$ we also have $\{v_8\} \subset \NH_G(v_4) \cap \NH_G(v_6) \cap \NH_G(v_7)$, therefore
    $\{v_4, v_6, v_7, v_8\}$ also induces a $G_{\tailtriangle}$.
    Since the information for $\Phi$ conveyed by this subgraph is also gathered from
    $\{v_2, v_4, v_5, v_6,v_7\}$ we do not consider it in the following.
    Since for $v_8$ there is no 3-clique containing it such that $v_5$ crosses only one chord of that 3-clique,
    there is no induced $G_{\tailtriangle}$ containing $v_5$ and $v_8$.
    Note that $\{v_2, v_4, v_5, v_8\}$ doesn't induce a $G_{\tailtriangle}$ since $\cauxx{v_2}{v_8}$ is not defined
    as $v_2$ and $v_8$ are neither on the same level nor does $v_2$ directly encase $v_8$.

    Since the chords in $\level{G}{}{3}$ do not encase any chords they do not induce any additional $G_{\tailtriangle}$.

    \subsection{Important Subgraph $G_{\boxslash}$}
    \label{subsubsec:imp2}

    Next, we consider the important subgraph $G_{\boxslash}$.
    This subgraph consists of two $3$-cliques sharing a common edge, resp.\ two vertices.
    The information we want to infer from this important subgraph for $\caux$ concerns the two non-adjacent
    vertices that are either in the same level or one of them directly encases the other.
    We consider each level and list all $G_{\boxslash}$ formed with chords in the adjacent level, resp.\ those that are
    directly encased, starting with the chords in $\level{G}{}{1}$.

\begin{figure}[tb]
        \centering
        \includegraphics{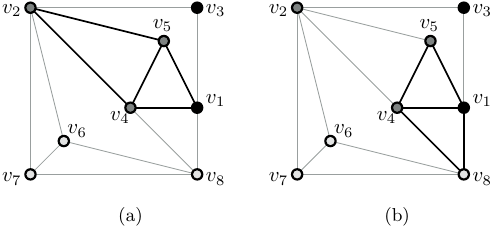}
        \caption{The different induced $G_{\boxslash}$.}
        \label{fig:counterex-boxslash}
    \end{figure}

    We have that $v_1$ directly encases $v_2$ and induces a $G_{\boxslash}$ with
    $\{v_4, v_5\} \subset \NH_G(v_1) \cap \NH_G(v_2)$, see Figure~\ref{fig:counterex-boxslash}(a).
    Since $\NH_G(v_1) \cap \NH_G(v_6) = \NH_G(v_1) \cap \NH_G(v_7) = \{v_8\}$, there is no induced $G_{\boxslash}$
    that contains $v_1$ and $v_6$, resp.\ $v_7$.
    Since $v_3$ is not part of any $3$-clique, the is also no induced $G_{\boxslash}$ containing $v_3$.

    For $\level{G}{}{2}$ we have that its chords form a clique, hence there is no induced $G_{\boxslash}$ with
    non-crossing chords in this level.
    Since $v_2$ does not encase any chords, it is not contained in any induced $G_{\boxslash}$.
    For $v_4$ we see that $\NH_G(v_4) \cap \NH_G(v_6) = \NH_G(v_4) \cap \NH_G(v_7) = \{v_2, v_8\} \notin E(G)$ and
    therefore there is no induced $G_{\boxslash}$ containing $v_4$ and $v_6$, resp.\ $v_7$.
    A similar argument holds for $v_5$ and $v_6$, resp.\ $v_7$, namely
    $\NH_G(v_5) \cap \NH_G(v_6) = \NH_G(v_5) \cap \NH_G(v_7) = \{v_2\}$, therefore $v_5$ also doesn't induce a $G_{\boxslash}$
    with $v_6$, resp.\ $v_7$.
    For $v_8$ however we have $\NH_G(v_5) \cap \NH_G(v_8) = \{v_1, v_4\} \in E(G)$, therefore $\{v_1, v_4, v_5, v_8\}$
    induces a $G_{\boxslash}$, see Figure~\ref{fig:counterex-boxslash}(b).

    The chords in $\level{G}{}{3}$ do not induce any $G_{\boxslash}$.

    \subsection{Important Subgraph $G_{\square}$}
    \label{subsubsec:imp3}

    The next important subgraph we consider is $G_{\square}$.
    For this important subgraph we list all chordless $4$-cycles, which are fairly easy to obtain from
    Figure~\ref{fig:counterexample}, and then show which of them form a $G_{\square}$, i.e.\ which of them give us
    information about two non-crossing chords in the same level or where one chord directly encases the other.

\begin{figure}[tb]
        \centering
        \includegraphics{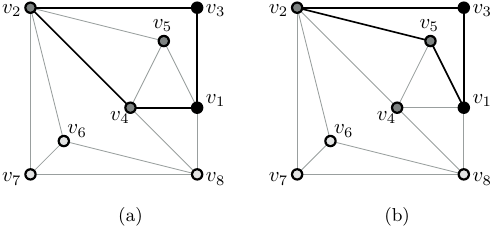}
        \caption{The different induced $G_{\square}$.}
        \label{fig:counterex-square}
    \end{figure}

    The chordless cycle of length $4$ are induced by
    $\{v_1, v_2, v_3, v_4\}$, $\{v_1, v_2, v_3, v_5\}$, $\{v_2, v_4, v_6, v_8\}$ and $\{v_2, v_4, v_7, v_8\}$.
    Since $v_1$ directly encases $v_2$ and $v_3$ directly encases $\{v_4, v_5\}$,
    $\{v_1, v_2, v_3, v_4\}$ and $\{v_1, v_2, v_3, v_5\}$ each form a~$G_{\square}$, see
    Figure~\ref{fig:counterex-square}(a), resp.~(b).
    Note that $v_2$ does not directly encase $v_8$ and they also do not belong to the same level, hence
    $\{v_2, v_4, v_6, v_8\}$ and $\{v_2, v_4, v_7, v_8\}$ induce no $G_{\square}$.

    \subsection{Important Subgraphs $G_{\pentagon}$ and $G_{\bigcirc}$}
    \label{subsubsec:imp4}

    The important subgraphs $G_{\pentagon}$ and $G_{\bigcirc}$ are not present in $G$.
    To see this, consider the chordless $4$-cycles in $G$.
    To extend any of them to a chordless cycle $C$ of length 5 or higher we would have to include both
    $v_1$ and $v_8$, therefore there is no $C$ such that
    $V(C) \subseteq \level{G}{}{i} \cup \level{G}{}{i+1}$ for any $i \in \{1,2\}$.

    \section{Counterexample to the Version of Unger's Thesis}
    \label{sec:second-counter}

\begin{figure}
        \centering
        \includegraphics[width=\textwidth]{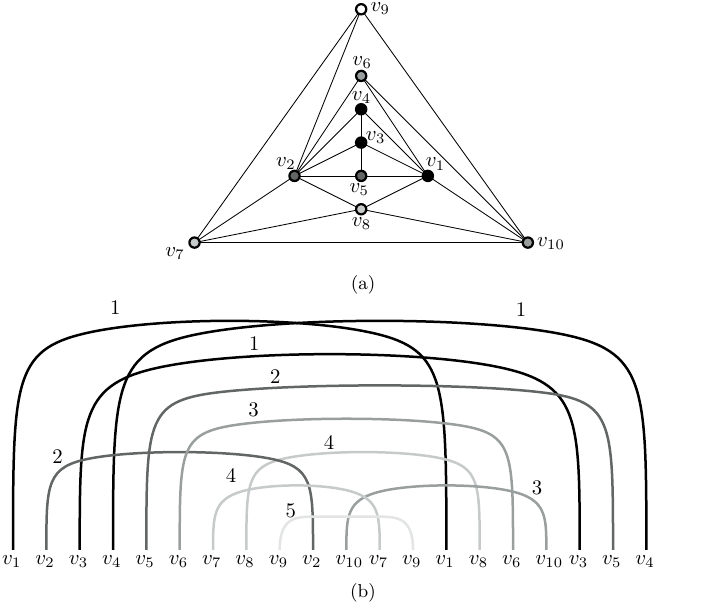}
        \caption{Counterexample for the thesis version of important subgraphs.}
        \label{fig:counterexample-diss}
    \end{figure}

    In the thesis~\cite[p.78]{DBLP:books/daglib/0070152}, the important subgraphs each are given their own specific
    definitions.
    The graph $G$ in Figure~\ref{fig:counterexample-diss} refutes the
    claim from Unger's thesis that states that a circle graph is 3-colorable if and only if $\Phi$ is satisfiable
    using the definitions from~\cite{DBLP:books/daglib/0070152}.

    The graph is clearly not 3-colorable, since in any 3-coloring $\col$ we have $\col(v_1) = \col(v_2)$ and
    $\col(v_1) = \col(v_7)$ due to the subgraphs induced by $\{v_1, v_2, v_3, v_4\}$ and $\{v_1, v_7, v_8, v_{10}\}$.
    However, $v_2$ and $v_7$ are adjacent.
    In the following we also show that there exists a satisfiable formula $\Phi(\mathcal{G})$ for the family of
    subgraphs $\mathcal{G}$ of $G$.

    \subsection{Important Subgraph $G_{\tailtriangle}$}
    \label{subsec:imp1}

    For the important subgraph $G_{\tailtriangle}$, the definition states that the non-adjacent vertices either
    belong to the same level or to two adjacent ones.
    We first consider those $G_{\tailtriangle}$ for which the non-adjacent vertices belong to one level.
    Since each level contains only three chords or less, there can be no
    $G_{\tailtriangle}$ formed by chords that are all on the same level.
    Therefore, we consider $G_{\tailtriangle}$ with non-adjacent vertices on two levels.
    For $\level{G}{}{1}$ and $\level{G}{}{2}$, it is easy to verify that no subset of $\{v_1, v_2, v_3, v_4, v_5\}$
    induces a $G_{\tailtriangle}$ for which the non-adjacent vertices lying in $\level{G}{}{1} \cup \level{G}{}{2}$.
    For $\level{G}{}{2}$ and $\level{G}{}{3}$ we have that the vertices $\{v_3, v_5, v_1, v_6, v_{10}\}$ induce four
    different $G_{\tailtriangle}$.
    The vertices $\{v_3, v_5, v_2, v_6\}$ also induce a $G_{\tailtriangle}$, although the clauses contributed by this
    important subgraph are already included in the aforementioned induced $G_{\tailtriangle}$.

    For $\level{G}{}{3}$ and $\level{G}{}{4}$ we have that $\{v_2, v_6, v_7, v_8\}$ and
    $\{v_6, v_7, v_8, v_{10}\}$ each induce a $G_{\tailtriangle}$ conveying the same information, namely $v_6$ must have
    the same color as either $v_7$ or $v_8$.
    Lastly, for $\level{G}{}{4}$ and $\level{G}{}{5}$ we have that no subset of these vertices induce a $G_{\tailtriangle}$.

    \subsection{Important Subgraph $G_{\boxslash}$}
    \label{subsec:imp2}

    Next, we consider the important subgraph $G_{\boxslash}$.
    These are defined such that the two non-adjacent vertices $u, v$ either directly encase each other or both endpoints
    of the chord corresponding to $u$ lie to the left (resp.\ right) of both endpoints of the chord corresponding to $v$.
    The pairs of non-adjacent vertices that belong to (at least) one induced $G_{\boxslash}$ are $\{v_1, v_2\}$,
    $\{v_4, v_5\}$, $\{v_1, v_7\}$, $\{v_2, v_{10}\}$, $\{v_6, v_8\}$ and $\{v_8, v_9\}$.
    Each of these pairs contributes one clause to $\Phi$.

    \subsection{Important Subgraph $G_{\square}$}
    \label{subsec:imp3}

    Observe, that every chordless $4$-cycle contains the non-adjacent pairs of vertices $\{v_1, v_2\}$ or
    $\{v_2, v_{10}\}$.
    Since these two pairs are already represented by the clauses contributed by $G_{\boxslash}$, which have to be set to
    \texttt{true}, every clause contributed by an induced $G_{\square}$ is already satisfied.

    \subsection{Important Subgraphs $G_{\pentagon}$ and $G_{\bigcirc}$}
    \label{subsec:imp4}

    The important subgraphs $G_{\pentagon}$ and $G_{\bigcirc}$ are not present in $G$, since there are no induced
    cycles of length~$> 4$.

    Tables~\ref{tab:imp-sg-diss} show the clauses we obtain from the induced important subgraph while Table~\ref{tab:counter2sol}
    gives a satisfying truth assignment for $\Phi$.

    \begin{table}[ht]
        \caption{Important subgraphs and the corresponding clauses for each of~(a)~$G_{\tailtriangle}$, (b) $G_{\boxslash}$, and
            (c) $G_{\square}$. For the satisfying variable assignment from Table~\ref{tab:counter2sol} all true literals are underlined.}
        \begin{center}
        \begin{subtable}{\textwidth}
            \centering
            \begin{tabular}{ | c | c | }
                \hline
                Important Subgraph $G_{\tailtriangle}$ & Clauses                                                                                              \\
                \hline
                $\{v_3, v_5, v_1, v_6\}$               & $(\underline{v_5 v_6} \vee v_3 v_6) \wedge (\neg                     v_5 v_6 \vee \underline{\neg v_3 v_6})$              \\
                \hline
                $\{v_3, v_5, v_1, v_{10}\}$            & $(\underline{v_3 v_{10}} \vee v_5 v_{10}) \wedge (\neg v_3 v_{10} \vee \underline{\neg v_5 v_{10}})$ \\
                \hline
                $\{v_3, v_1, v_6, v_{10}\}$            & $(\underline{v_3 v_{10}} \vee v_3 v_{6}) \wedge (\neg v_3 v_{10} \vee \underline{\neg v_3 v_{6}})$   \\
                \hline
                $\{v_5,  v_1, v_6, v_{10}\}$             & $(v_5 v_{10} \vee \underline{v_5 v_{6}}) \wedge (\underline{\neg v_5v_{10}} \vee \neg v_5 v_{6})$    \\
                \hline
                $\{v_2, v_6, v_7, v_8\}$               & $(\underline{v_6 v_8} \vee v_6 v_7) \wedge (\neg v_6 v_8 \vee \underline{\neg v_6 v_7})$             \\
                \hline
                $\{v_6, v_7, v_8, v_{10}\}$            & $(\underline{v_6 v_8} \vee v_6 v_7) \wedge (\neg v_6 v_8\vee \underline{\neg v_6 v_7})$              \\
                \hline
            \end{tabular}
            \caption{}
        \end{subtable}
        \begin{subtable}{\textwidth}
            \centering
            \begin{tabular}{ | c | c | }
                \hline
                Important Subgraph $G_{\boxslash}$           & Clauses               \\
                \hline
                $\{v_1, v_2, v_3, v_i\}$ for $i \in \{4,5\}$ & \underline{$v_1 v_2$} \\
                \hline
                $\{v_1, v_{7}, v_i, v_j\}$ for $i \in \{2,8\}$, $j \in \{8,10\}$, $i \neq j$ & \underline{$
                v_2 v_{7}$} \\
                \hline
                $\{v_2, v_{10}, v_7, v_i\}$ for $i \in \{8,9\}$ & \underline{$
                v_2 v_{10}$} \\
                \hline
                $\{v_4, v_{5}, v_3, v_i\}$ for $i \in \{1,2\}$ & \underline{$
                v_4 v_{5}$} \\
                \hline
                $\{v_6, v_{8}, v_1, v_{10}\}$ & \underline{$
                v_6 v_{8}$} \\
                \hline
                $\{v_8, v_{9}, v_7, v_i\}$ for $i \in \{2,10\}$ & \underline{$
                v_8 v_{9}$} \\
                \hline
            \end{tabular}
            \caption{}
        \end{subtable}
        \begin{subtable}{\textwidth}
            \centering
            \begin{tabular}{ | c | c | }
                \hline
                Important Subgraph $G_{\square}$                                              & Clauses                              \\
                \hline
                $\{v_1, v_2, v_i, v_j\}$ for $i \in \{3,4,5\}$, $j \in \{5,6,8\}$, $i \neq j$ & \underline{$v_1 v_2$} $\vee v_i v_j$ \\
                \hline
                $\{v_2, v_{10}, v_i, v_j\}$ for $i \in \{6,8\}$, $j \in \{7,8,9\}$, $i \neq j$ & $\underline{v_2 v_{
                    10}}$ $\vee v_i v_j$ \\
                \hline
            \end{tabular}
            \caption{}
        \end{subtable}

        \hfill
        \label{tab:imp-sg-diss}
        \end{center}
    \end{table}

    \begin{table}
        \caption{Satisfying truth assignment for the second counterexample.}
        \begin{center}
            \begin{tabular}{ | c | c | }
                \hline
                True & $v_3 v_{10}$, $v_5 v_6$, $v_6 v_8$, $v_7 v_8$, $v_9 v_{10}$, $v_2 v_{11}$, $v_1 v_2$, $v_4
                v_5$, $
                v_1 v_7$ \\
                \hline
                False & $v_3 v_6$,$v_5 v_{10}$, $v_6 v_7$ \\
                \hline
            \end{tabular}
        \end{center}
        \label{tab:counter2sol}
    \end{table}

    \section{Counterexample to Unger's Backtracking Algorithm}
    \label{sec:bt-counter}

\begin{figure}
        \centering
        \includegraphics{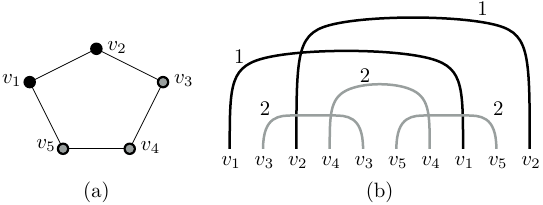}
        \caption{Graph for which Unger's backtracking might return the wrong result.}
        \label{fig:bt-counter}
    \end{figure}

    Consider a cycle of length 5 and a corresponding chord diagram; see Fig.~\ref{fig:bt-counter}.
    We show that the backtracking algorithm described by Unger can be executed such that no satisfying variable
    assignment is found, even though the graph is clearly $3$-colorable.
    In this case the only important subgraph is the graph itself.
    We now give the clauses constructed for important subgraphs that are isomorphic to induced cycles as
    described in Unger's thesis~\cite[p.~87]{DBLP:books/daglib/0070152}.
    More generally, let $G = (V, E)$ with $V = \{v_1,\dots, v_{k}\}$ be a circle graph isomorphic to a cycle $C_k$ with
    $k \geq 5$ and let $\caux$ be an auxiliary coloring function for $G$ with
    respect to $\mathcal{P}(G)$.
    The function $h$ is recursively defined as follows:
    \[
        \begin{array}{ll}
            h(v_x, v_{x+1}) \coloneqq \texttt{false} \text{ for } x \in \{1,2\} & \text{ and
            } \\
            h(v_x, v_{x+2}) \coloneqq \caux(v_x, v_{x+2}) \text{ for } x \in \{1,2\} & \text{ and }                                    \\
            h(v_x, v_i) = \left\{
            \begin{array}{ll}
                h(v_x, v_{i-2})                                  & \text{ if } \caux(v_{i-2}, v_i) = \texttt{true} \\
                \neg h(v_x, v_{i-2}) \wedge \neg h(v_x, v_{i-1}) & \text{ if } \caux(v_{i-2},v_i) = \texttt{false}
            \end{array}
            \right.
        \end{array}
    \]
    for $x \in \{1,2\}$ and $x+3 \leq i \leq k$.
    Unger shows that $\caux$ is realizable for $G$ if and only if all of the following conditions hold:
    \begin{itemize}
        \item $h(v_1, v_{k}) = \texttt{false}$
        \item $h(v_1, v_{k-1}) = \caux(v_1, v_{k-1})$
        \item $h(v_2, v_k) = \caux(v_2, v_k)$.
    \end{itemize}
    From this we infer the following formula $\Phi(G)$ for $G$ isomorphic to $C_5$:
    \[
        \begin{array}{ll}
            \underline{\neg h(v_1, v_2)} \wedge \underline{\neg h(v_2,v_3)} \wedge \underline{\neg h(v_1,v_5)}                    & \wedge \\
            (h(v_1, v_{3}) \vee \neg (v_1, v_3)) \wedge (\neg h(v_1, v_3) \vee (v_1, v_3))                                                           & \wedge \\
            (h(v_2, v_{4}) \vee \neg (v_2, v_4)) \wedge \underline{(\neg h(v_2, v_{4}) \vee (v_2, v_4)})
            & \wedge
            \\
            (h(v_1, v_{4}) \vee \neg (v_1, v_{4})) \wedge (\neg h(v_1, v_{4}) \vee (v_1, v_{4}))
            & \wedge
            \\
            (h(v_2, v_5) \vee \neg (v_2, v_5)) \wedge (\neg h(v_2, v_5) \vee (v_2, v_5))                                                             & \wedge \\
            \underline{(\neg (v_{i-2}, v_i) \vee \neg h(v_1, v_i) \vee h(v_1, v_{i-2}))} \wedge \underline{(\neg (v_{i-2}, v_i) \vee h(v_1, v_i) \vee \neg h(v_1, v_{i-2}))}
            & \wedge
            \\
            \underline{(\neg (v_{j-2}, v_j) \vee \neg h(v_2, v_j) \vee h(v_2, v_{j-2}))} \wedge (\neg (v_{j-2}, v_j) \vee h(v_2,
            v_j) \vee \neg h(v_2, v_{j-2}))
            & \wedge
            \\
            ((v_{i-2}, v_i) \vee \neg h(v_1, v_i) \vee \neg h(v_1, v_{i-2})) \wedge \underline{((v_{i-2}, v_i) \vee \neg h(v_1, v_i) \vee \neg h(v_1, v_{i-1}))}
            & \wedge
            \\
            ((v_{i-2}, v_i) \vee h(v_1, v_i) \vee h(v_1, v_{i-2}) \vee h(v_1, v_{i-1}))                                                              & \wedge \\
            ((v_{j-2}, v_j) \vee \neg h(v_2, v_j) \vee \neg h(v_2, v_{j-2})) \wedge \underline{((v_{j-2}, v_j) \vee \neg h(v_2,v_j) \vee \neg h(v_2, v_{j-1}))}
            & \wedge
            \\
            \underline{((v_{j-2}, v_j) \vee h(v_2, v_j) \vee h(v_2, v_{j-2}) \vee h(v_2, v_{j-1}))}                                                              &        \\
        \end{array}
    \]
    with $i \in \{3,4,5\}$ and $j \in \{4,5\}$.
    The clauses that are relevant to our counterexample are underlined.
    Since Unger does not specify an order in which the variables are assigned, we may assign values to a subset of
    variables in the following order:
    \begin{gather*}
        h(v_2, v_5), h(v_1, v_4), h(v_2, v_4), h(v_1, v_3), h(v_1, v_5), h(v_2, v_3), \\
        h(v_1,v_2), (v_3, v_5), (v_2, v_4)\\
    \end{gather*}
    We apply the following strategy for jumping in the backtracking tree, which is also consistent with Unger's
    description of his backtracking algorithm.
    Recall, that the vertices of the backtracking tree represent variables and their child edges
    correspond to their truth assignments.
    Further, when encountering a clause whose literals are all set to \texttt{False} during
    the variable assignment, we may jump in the backtracking
    tree to a variable of that clause
    that (1) results in a literal of that clause being set to \texttt{True} by flipping
    its assigned value and (2) the child edge that corresponds to this new assigned value does not exist,
    i.e.\ we haven't tried both truth assignments at that vertex yet.
    For each variable we first set its value to \texttt{True} and check whether this assignment causes a
    clause to contain only \texttt{False} literals.
    If so, we flip the truth assignment of that variable to \texttt{False} and check, if
    any other clause now contains only \texttt{False} literals.
    If there still exists a clause whose literals are all set to \texttt{False},
    we may pick a variable from that respective clause that lies on the path from the current variable to
    the root and is not currently set to \texttt{False}.
    If no such variable exists, there is no variable that satisfies properties (1) and (2) of Unger's
    description, and therefore the algorithm reports that the graph is not $3$-colorable.

\begin{figure}[t]
        \centering
        \includegraphics[width=\textwidth]{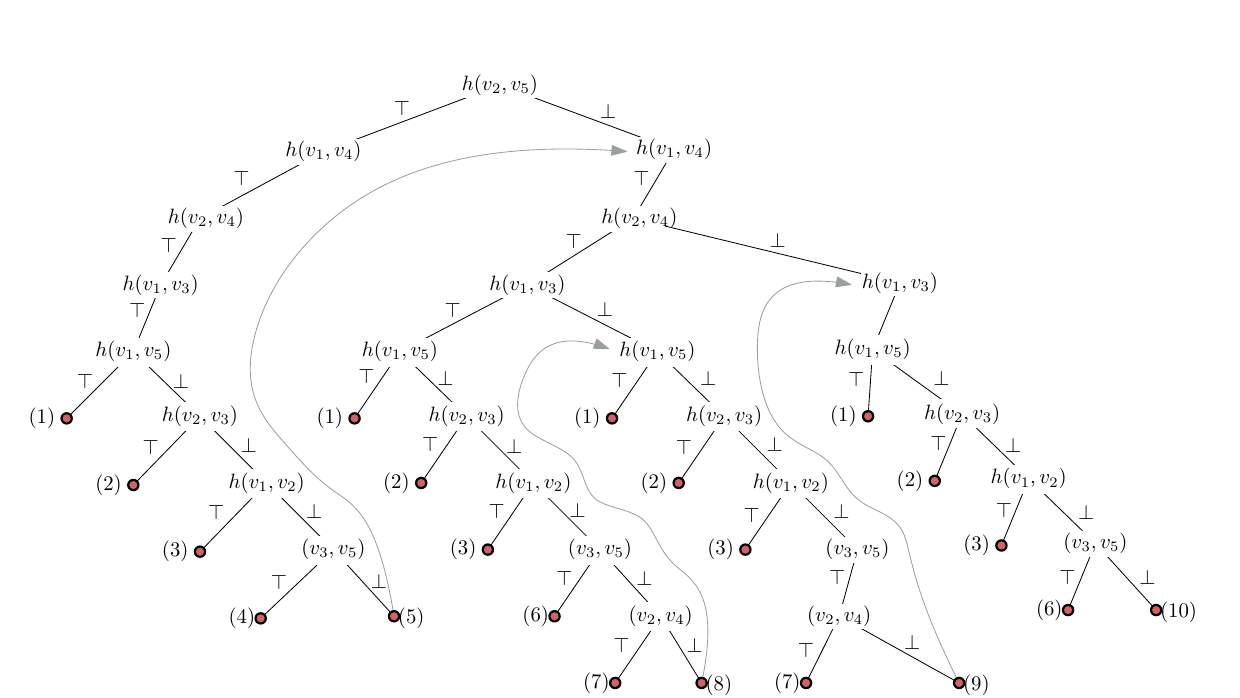}
        \caption{Backtracking tree for $C_5$.}
        \label{fig:bt-tree}
    \end{figure}

    We apply this strategy to the variables listed above.
    The resulting backtracking tree is illustrated in Figure~\ref{fig:bt-tree}.
    The contradicting clauses and jumps in the backtracking tree that occur are shown in Table~\ref{tab:bt}.
    \begin{table}[ht]
        \caption{Unsatisfied clauses and jumps performed during backtracking.}
        \begin{center}
            \begin{tabular}{ | c | c | c |}
                \hline
                Pos.\ in backtracking tree & Contradicting clause                                            & Jumping to value \\
                \hline
                (1)                        & $\neg h(v_1, v_5)$                                              & $h(v_1, v_5)$    \\
                \hline
                (2)                        & $\neg h(v_2, v_3)$                                              & $h(v_2, v_3)$    \\
                \hline
                (3)                        & $\neg h(v_1, v_2)$                                              & $h(v_1, v_2)$    \\
                \hline
                (4)                        & $\neg(v_3, v_5) \vee \neg h(v_2, v_5) \vee h(v_2, v_3)$         & $h(v_3, v_5)$    \\
                \hline
                (5)                        & $(v_3, v_5) \vee \neg h(v_2, v_5) \vee \neg h(v_2, v_4)$        & $h(v_2, v_5)$    \\
                \hline
                (6)                        & $\neg (v_3, v_5) \vee \neg h(v_1, v_3) \vee h(v_1, v_5)$        & $(v_3, v_5)$     \\
                \hline
                (7)                        & $\neg (v_2, v_4) \vee \neg h(v_1, v_4) \vee h(v_1, v_2)$        & $(v_2, v_4)$     \\
                \hline
                (8)                        & $(v_2, v_4) \vee \neg h(v_1, v_3) \vee \neg h(v_1, v_4)$        & $h(v_1, v_3)$    \\
                \hline
                (9)                        & $(v_2, v_4) \vee \neg h(v_2, v_4)$                              & $h(v_2, v_4)$    \\
                \hline
                (10)                       & $(v_3, v_5) \vee h(v_2, v_5) \vee h(v_2, v_3) \vee h(v_1, v_4)$ & n.a.             \\
                \hline
            \end{tabular}
            \label{tab:bt}
        \end{center}
    \end{table}
    For this specific order of variables, the algorithm runs out of variables to flip in clause (10) and therfore
    incorrectly reports that $C_5$ is not 3-colorable.

    \section{Test Data}
    \label{sec:test-data}

    We evaluated Unger's backtracking algorithm on 2196 graphs ranging from 8 to 750 vertices.
    We generate 3-colored circle graphs by picking a random number within the range of 3 to 750 as the
    desired number of vertices and then inserting two endpoints of a chord representing vertex
    $v_i$ into two randomly chosen distinct cells of an array.
    If after inserting the chord $v_i$ the graph does not contain a clique of size greater than $3$ and the
    current coloring can be extended to $v_i$, i.e.\ there is at least
    one color left that is not yet used by any neighbor of $v_i$, $v_i$ is kept.
    Otherwise it is discarded, and we repeat these steps for a new pair of randomly inserted endpoints.
    This is done until the desired number of vertices has been reached.

\end{document}